\documentclass[twocolumn,showpacs,preprintnumbers,amsmath,amssymb,floatfix,prb,a4paper]{revtex4}
\usepackage[colorlinks,hypertex]{hyperref}

\usepackage{times}
\usepackage{epsfig}

\newcommand{\be}[1]{\begin{equation}\label{#1}}
\newcommand{\ee}{\end{equation}}   
\newcommand{\eq}[1]{(\ref{#1})}

\newcommand{\showlabel}[1]{}

\begin{document}

\title{
	Coarse Grained Simulations of a Small Peptide: Effects of Finite 
	Damping and Hydrodynamic Interactions
}

\author{Uwe Winter}
\author{Tiham\'{e}r Geyer}
\email[Author to whom correspondence should be addressed. Electronic 
	   mail: ]{Tihamer.Geyer@bioinformatik.uni-saarland.de}
\affiliation{Zentrum f\"ur Bioinformatik, Universit\"at des 
Saarlandes, D--66123 Saarbr\"ucken, Germany}


\begin{abstract}
	In the coarse grained Brownian Dynamics simulation method the many
	solvent molecules are replaced by random thermal kicks and an
	effective friction acting on the particles of interest. For
	Brownian Dynamics the friction has to be so strong that the
	particles' velocities are damped much faster than the duration of
	an integration timestep. Here we show that this conceptual limit
	can be dropped with an analytic integration of the equations of
	damped motion. In the resulting Langevin integration scheme our
	recently proposed approximate form of the hydrodynamic
	interactions between the particles can be incorparated
	conveniently, leading to a fast multi-particle propagation scheme,
	which captures more of the short-time and short-range solvent
	effects than standard BD. Comparing the dynamics of a bead-spring
	model of a short peptide, we recommend to run simulations of small
	biological molecules with the Langevin type finite damping and to
	include the hydrodynamic interactions.
\end{abstract}

%

\maketitle

\section{Introduction}

For the simulation of microscopic particles in a solvent, Brownian
Dynamics (BD) has become a work horse technique, which is routinely
applied to study, e.g., the properties of colloidal suspensions
\cite{PHI88,LAD90,GOR04}, the dynamics of polymers
\cite{LIU03,BEA00b,SZY07,EMP08}, or the association of biological proteins
\cite{GAB02,MIN05,SPA06,DLU08}. Einstein's seminal explanation of the
random thermal motion of microscopic objects in a solvent
\cite{EIN05}, which had been discovered earlier by Robert Brown, is
based on the assumption that the solvent molecules are so much smaller
than the interesting objects that they can be replaced by a continuous
solvent, which then modifies the motion of the much larger Brownian
particles in three major ways. First, the kinetic energy of the large
particles is dissipated via the viscosity of the solvent. Second, the
thermal motion of the solvent molecules is replaced by random kicks to
the observed particles. These two contributions effectively form a
thermostat. Third, the direct interactions like Coulomb or short range
van-der-Waals interactions are modified by the solvent. With these
three modifications, the problem of solving Newton's equations of
motion for very many particles with well-defined position-dependent
interaction potentials is replaced by solving a few-body problem with
more complex interactions, a velocity-dependent dissipation, and an
additional noise term. This level of approximation is often referred
to as Langevin Dynamics (LD), whereas Brownian Dynamics involves the
additional approximation that only "long" time intervals are
considered \cite{DHO96}. Then, due to the damping of the particle
velocities, all information about the momentum is lost between two
subsequent observations. This is called the overdamped regime.

Obviously, the central criterion whether BD is an adequate simulation
method, is that the solvent molecules are indeed much faster---or that
the Brownian particles are much larger and thus slower than a water
molecule with its effective diameter of about 1.5 \AA. For colloidal
particles with diameters on the micron scale this separation of scales
is a good approximation, but for the much smaller biological proteins
with diameters of a few nanometers it is already questionable. This
becomes even more pronounced, when this coarse grained method is
applied to the internal dynamics of proteins.

In addition to the thermostat effect, the solvent also mediates a
mechanical coupling between the observable particles, the so called
hydrodynamic interactions (HI). The resulting correlations in the
particle velocities may have striking effects on the folding behavior
of proteins as shown recently \cite{FRE09}.

Here, we investigate the effects of the approximations outlined above.
For a bead-spring representation of a short peptide we compare plain
BD simulations to BD simulations with HI included and also to LD
simulations both with and without HI to see how the simulated dynamics
of very small particles is influenced by the non-negligible relaxation
times and how the hydrodynamic coupling between the beads affects the
overall dynamics. In the next section we introduce our implementation
of the LD and BD algorithms and how hydrodynamic interactions can be
incorporated efficiently. There we also explain the coarse graining
procedure used to set up the bead-spring model of the peptide. The
results then show how the different approximations influence the
dynamic behavior of the model peptide.

\section{Methods}

\subsection{Langevin and Brownian Propagation}

For the following it is convenient to formulate the Brownian and the
Langevin propagation algorithms with effective forces instead of
directly using displacements as in the original BD scheme of Ermak and
McCammon \cite{ERM78}. For this, we start from Newton's equations of
motion for the complete system including all solvent molecules.
Obviously, the particle masses $m_{i}$ can be taken as constant such
that the change of the momenta $p_{i}$ can be expressed as a change of
the velocities $v_{i}$ due to the sums $F_{i}$ of all pairwise
position-dependent forces.
\be{eq:NewtonAll}
	\frac{d p_{i}}{d t} = m_{i}\frac{d v_{i}}{d t} = 
		F_{i} = \sum_{k \neq i} F(r_{ik})
\ee
The dimensionality of this system of equations can be greatly reduced
by assuming that the solvent molecules are very small. Then their
masses are very small and they move so fast that their effect on the
slower degrees of freedom of the interesting particles can be treated
as a mean-field heat bath consisting of a dissipative friction term
plus a random driving \cite{EIN05}. For a spherical particle, the
phenomenological friction constant $\gamma$ can be calculated via the
Stokes-Einstein relation from the hydrodynamic radius $a$ of the
particle and the macroscopic bulk viscosity $\eta$ of the solvent as
$\gamma = 6\pi\eta a$ \cite{DHO96}. Due to the random nature of the
driving kicks from the solvent molecules, these kicks are
characterized via the statistical measures of a vanishing average and
a finite covariance of the resulting displacements $R_{i}(\delta t)$:
\be{eq:Statistics}
	\langle R_{i} \rangle = 0 \quad \mbox{and} \quad
	\langle R_{i} R_{k} \rangle = 2 D_{ik} \delta t
\ee

For the diagonal terms with $i=k$ the diffusion coefficient $D_{ii} =
k_{B}T / \gamma_{i}$ gives the ratio between the solvent induced
thermal energy $k_{B}T$ and the also solvent mediated dissipation. For
the off-diagonal terms, $D_{ik}$ describes the hydrodynamic coupling
due to the displaced solvent. 

To determine the random displacements $R_{i}$ from the covariance
\eq{eq:Statistics}, one essentially has to take the square root of the
diffusion matrix \cite{FIX81}. Due to the high numerical cost associated
with this step, the hydrodynamic coupling was often neglected in BD
simulations by setting the off-diagonal entries $D_{ik} = 0$. It was
only recently that we could show how this expensive matrix
factorization can be approximated by a sum of two-body contributions
and the correlated random displacements can be expressed via effective
random forces \cite{GEY09} $f_{i}^{eff}$, leading to the same
$\mathcal{O}(N^2)$ runtime scaling as for the pairwise direct
interactions $F_{ik}$.

To arrive at a formulation of the equation of motion with the implicit
solvent which is convenient for numerical propagation, we start from
\eq{eq:NewtonAll} with the total force $F = F_{ext} + f$, which is the
sum of the external forces $F_{ext}$ and the random kicks $f$:
\be{eq:NewtonOne}
	\frac{dv}{dt} = \frac{1}{m} (F - \gamma v)
\ee
For simplicity, we omit the coordinate index $i$ for the following.
This equation of motion can be integrated analytically over one
timestep of length $\Delta t$, which must be so small that the
configuration dependent forces remain essentially constant
\cite{BEA00a}. Due to the linear Stokes friction, we can use the
average $f(\Delta t)$ of the random kicks. With the initial velocity
$v_{0}$, $v(\Delta t)$ at the end of the timestep is then
\be{eq:V1Step}
	v(\Delta t) = \frac{F}{\gamma} + 
	\left(v_{0} - \frac{F}{\gamma}\right) 
	\mbox{e}^{ - \frac{\gamma \Delta t}{m}}
\ee
This equation can be integrated once more to give the displacement
$\Delta x(\Delta t)$ during $\Delta t$:
\be{eq:DX1Step}
	\Delta x(\Delta t) = \frac{F}{\gamma}\Delta t -
		\frac{m}{\gamma}\left(\frac{F}{\gamma} - v_{0}\right)
		\left(1 - 
		\mbox{e}^{ - \frac{\gamma \Delta t}{m}} \right)
\ee

From these equations the conventional BD propagation algorithm is
obtained by going to the overdamped regime where $\Delta t \gg
\tau_{rel}$, i.e., only timesteps $\Delta t$ are considered that are
much longer than the velocity relaxation time $\tau_{rel} = m/\gamma$.
Then $v_{0}$ can be neglected and the displacement simplifies to
\be{eq:DXBrown}
	\Delta x(\Delta t) = \frac{F}{\gamma}\Delta t = 
	\frac{D_{0}}{k_{B}T} F\, \Delta t,
\ee
which can also be expressed with the diffusion constant $D_{0}=k_{B}T 
/ \gamma$ of the particle.

For the rotational motion, analogous equations are used where 
position $x$ and velocity $v$ are replaced by a rotation angle and 
an angular velocity, and torques instead of the forces $F$ act on 
moments of inertia.

\begin{figure}[t]
	\begin{center}
		\includegraphics[scale=0.7]{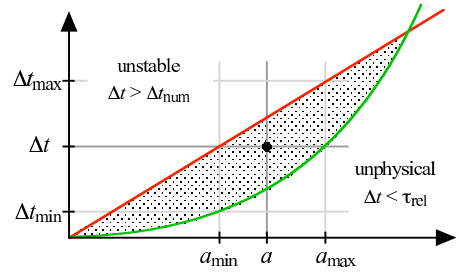}
	\end{center}
	\caption{Range of admissible timesteps $\Delta t$ between the
	inaccurate and the unphysical regimes of a BD simulation vs.
	particle radius $a$. The lower green line denotes the conceptual
	limit according to $\Delta t \gg \tau_{rel} \propto a^2$. The
	upper red limit indicates that the maximal timestep for a
	numerically stable propagation increases only $\propto a$.
	Consequently, for any $\Delta t$ there is a minimal particle size
	$a_{min}$, for which the propagation becomes unstable, and a
	maximal $a_{max}$, for which the assumption of overdamped motion
	is not valid anymore. For further explanations see text.}
	\label{fig:DTLimits}
\end{figure}

The Langevin equations \eq{eq:V1Step} and \eq{eq:DX1Step} look much
more complex than the Brownian propagator \eq{eq:DXBrown}, but for
practical applications the main effort is to determine the forces and
torques acting on each of the particles. Thus, keeping track of the
velocity only incurs a negligibly small additional overhead for the
benefit that one has not to make sure that the conceptual constraint
of BD, $\Delta t \gg \tau_{rel}$, is fulfilled. It is actually quite
easy to find simulation scenarios where these BD requirements cannot
be fulfilled for all particles simultaneously. This can be seen from
the following estimate of the two limits for $\Delta t$. The first is
that $\Delta t \gg \tau_{rel} = m/\gamma$. For typical colloidal
particles or biological polymers, the mass $m = \frac{4\pi}{3} \rho
a^3$ is proportional to the volume and thus scales $\propto a^3$ with
the radius $a$. From the Stokes-Einstein relation we get $\gamma = 6
\pi \eta a \propto a$, thus $\tau_{rel} = \frac{2}{9} \frac{\rho
a^2}{\eta}$ increases quadratically with $a$. On the other hand,
$\Delta t$ has to be small enough for numerical reasons to keep the
typical displacements $\Delta x(\Delta t) \ll \Delta x_{max}$, where
$\Delta x_{max}$ is a typical potential well extensions or the
dimensions of the smallest particles. According to \eq{eq:DXBrown}
this upper limit scales $\propto a$. In a simulation where all
particles have the same size, usually a $\Delta t$ can be found, which
fulfills both requirements. However, when particles of different sizes
are considered, a timestep which allows for a stable propagation of
the small particles may be much shorter than $\tau_{rel}$ for the
largest particles, while a $\Delta t$ that ensures the overdamped
regime for the larger particles most likely leads to numerical
instabilities. This problem is sketched in figure \ref{fig:DTLimits},
where the green line denotes the lower limit from the overdamped
regime and the red line the upper numerical stability margin. The
dotted area between these two limits is the area of permissible
timesteps. For a fixed $a$ there is a certain region $[\Delta t_{min},
\Delta t_{max}]$ for $\Delta t$ to choose from, but when particles of
different sizes are used in the same simulation, then for a given
timestep $\Delta t$ there is only a range of particle sizes $[a_{min},
a_{max}]$ which can be propagated reliably within the assumption of
Brownian Dynamics.

A few numbers highlight this problem. With a typical protein density
of $\rho = 1.1 \rho_{H_{2}O}$ and $\rho / \eta = 1.23$ ps/nm$^2$ we
find that a protein of $a$ = 5 nm radius has $\tau_{rel}= 6.8$ ps. The
small electron carrier cytochrome $c_{2}$ with $a = 1.67$ nm has only
$\tau_{rel} = 0.76$ ps and an ion with an effective hydrodynamic
radius of 0.2 nm looses its velocity already within $\tau_{rel} =
0.01$ ps. In a simulation with cytochrome $c_{2}$ and ions, based on
practical experience, a timestep of $\Delta t = 1$ ps still yields a
stable propagation of the fast ions. While the ions are clearly in the
overdamped regime, this assumption is at least questionable for the
cytochrome $c_{2}$. When one wants to simulate the association of
cytochrome $c_{2}$ to larger proteins in the presence of explicitly
modelled ions, $\Delta t = 1$ ps is clearly too short for a
theoretically valid overdamped BD description.

We note that for the analytical integration above we required that the
interparticle forces be constant during one timestep. When the
underlying interaction potentials are not approximated by constant
slopes but one order further by harmonic potentials, the analytical
solutions for the motion of a damped harmonic oscillator can be used
for the propagator. Though this is not really practical, it allows to
distinguish between the strongly damped creeping and the weakly damped
ballistic regimes based on the ratio of potential curvature and
relaxation time. With this criterion and the parameter values above,
only for very short ranged potentials like van-der-Waals interactions
the dynamics may come close to the weakly damped ballistic and thus
oscillatory regime.

\subsection{Including Hydrodynamic Interactions}

When one wants to include HI into a Langevin propagation as the one
from equations \eq{eq:V1Step} and \eq{eq:DX1Step}, then there is a
conceptual difficulty. The usual Oseen \cite{KIR48} or
Rotne-Prager-Yamakawa \cite{ROT69,YAM70} (RPY) hydrodynamic tensors
are built with the help of the Fax\'en theorem \cite{FAX22} from the
stationary flow fields around spheres moving with constant velocities.
In an overdamped BD simulation one can at least assume that the
velocities are constant during one timestep and then treat each
timestep as a configuration with (temporarily) stationary velocities
mimicking these creeping flow conditions. Due to the linear Stokes
friction, the velocity does not even occur in the BD propagator
\eq{eq:DXBrown} and the (constant) forces can directly be converted
into displacements.

With hydrodynamic interactions the BD propagator \eq{eq:DXBrown} for
the $i$th coordinate becomes \cite{ERM78}
\be{eq:BDWithHI}
	\Delta x_{i}(\Delta t) = 
		\sum_{k} \frac{D_{ik} F_{k}}{k_{B}T}  \Delta t + 
		R_{i}(\Delta t),
\ee
which can be rewritten with a hydrodynamically corrected effective 
force \cite{GEY09} $F_{i}^{eff} = \sum_{k}\frac{D_{ik}}{D_{ii}}F_{k}$:
\be{eq:EffForce}
	\Delta x_{i}(\Delta t) = 
		\frac{D_{ii}\,\Delta t}{k_{B}T} F_{i}^{eff}  + R_{i}(\Delta t)
\ee
With the RPY HI tensor, the term taking care of local variations of
the diffusion coefficients vanishes and is omitted here (see, e.g.,
the original derivation by Ermak and McCammon \cite{ERM78}). Recently,
we could show \cite{GEY09} how the expensive calculation of the
correlated random displacements $R_{i}(\Delta t)$ can be approximated
efficiently with an ansatz that converts the uncorrelated random
forces $f_{i}$ into hydrodynamically corrected effective random forces
$f_{i}^{eff}$:
\be{eq:HIAnsatz}
	R_{i}(\Delta t) = \frac{D_{ii}\Delta t}{k_{B}T} C_{i}
		\sum_{k} \beta_{ik} \frac{D_{ik}}{D_{ii}} f_{i}
		= \frac{D_{ii} \Delta t}{k_{B}T} f_{i}^{eff}
\ee
With $\beta_{ii} = 1$, the normalization factors can be determined 
from
\be{eq:Ci}
	\left( \frac{1}{C_{i}}\right)^2
	= 1 + \sum_{k \neq i} \beta_{ik}^2 \frac{D_{ik}^2}{D_{ii}D_{kk}}
\ee
and the quadratic equation
\be{eq:betaik}
	\beta_{ik} = \frac{1 - \sqrt{1 - [(N-1)\epsilon^2 - (N-2)\epsilon]}}
		{(N-1)\epsilon^2 - (N-2)\epsilon} ,
\ee
where $\epsilon = \langle D_{ik} / D_{ii} \rangle$ is the average
normalized off-diagonal entry of the diffusion matrix. Then, the
displacements $\Delta x_{i}$ can be calculated efficiently with an
overall $\mathcal{O}(N^2)$ runtime scaling from the hydrodynamically
corrected external and random forces and the diagonal entries of the
diffusion matrix as
\be{eq:BDeffForces}
	\Delta x_{i}(\Delta t) = \frac{D_{ii}\, \Delta t}{k_{B}T} 
		(F_{i}^{eff} + f_{i}^{eff})
\ee

We now show how this idea can also be used in an LD description, where
the velocities are not constant during a timestep but change from the
initial $v_{0}$ to the final $v(\Delta t)$, which is then the new
$v_{0}$ for the next step. For this, we need an effective force
$\tilde{F}$ which is constant during one timestep and leads to the
same total displacement as with the correct propagator. With 
$\Delta x(\Delta t) = \tilde{F} \Delta t / \gamma$ we find from 
equation \eq{eq:DX1Step}
\be{eq:FTilde}
	\tilde{F} = \left[1 - \frac{m}{\gamma \Delta t} \left(
		1 - \mbox{e}^{ - \frac{\gamma \Delta t}{m}} \right)
		\right] F
		+ \frac{m}{\Delta t}\left(
		1 - \mbox{e}^{ - \frac{\gamma \Delta t}{m}} \right) 
		v_{0}.
\ee
The new $v_{0}$ for the next timestep is then
\be{eq:V0Tilde}
	\tilde{v}_{0} = \exp\left[ - \frac{\gamma \Delta t}{m}\right] v_{0}
		+ \frac{1}{\gamma} \left( 
			1 - \mbox{e}^{ - \frac{\gamma \Delta t}{m}}
		\right) F\, .
\ee
In the actual implementation equations \eq{eq:FTilde} and
\eq{eq:V0Tilde} occur twice, once for the external forces $F_{i}$ and
once for the random forces $f_{i}$. From these damping-corrected
uncorrelated forces the hydrodynamically correlated forces
$\tilde{F}_i^{eff}$ and $\tilde{f}_{i}^{eff}$ are then computed as
explained above and used to propagate  the particles according to
equation \eq{eq:BDeffForces}. With this recipe, our recent efficient
truncated expansion HI \cite{GEY09} can be combined with a Langevin
type propagation of Brownian particles, which is conceptually valid
even for arbitrarily small timesteps with only negligibly more effort.
Most of the prefactors in equations \eq{eq:FTilde} and \eq{eq:V0Tilde}
need only be evaluated once during the simulation setup.

\section{Results}

As a testcase for the propagation schemes outlined above we used a
small eleven-residue peptide, which had been used in our group in a recent
molecular dynamics (MD) study of the effects of interfacial water
layers during peptide-protein association at an SH3 domain
\cite{AHM08}. The sequence of the peptide is SHRPPPPGHRV using the
one-letter codes \cite{ALB02}. The four central prolin (P) residues
form a rather rigid PP2 helix, the short peptide therefore does not
show any folding or unfolding on the considered timescales.

\begin{figure}[t]
	\begin{center}
		\includegraphics[scale=0.9]{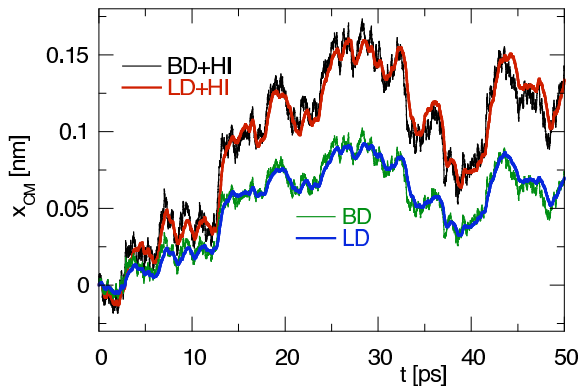}
	\end{center}
	\caption{X-component of the trajectory of the peptide CM under the
	same sequence of random kicks with the different propagation
	schemes. The inclusion of HI accelerates the CM diffusion and the
	finite damping of LD smoothes and slightly delays the peptide's
	response to the fast thermal kicks without changing the long-time
	trajectory.}
	\label{fig:Trajs}
\end{figure}

The atomic structure of the peptide was coarse-grained by first
placing a van-der-Waals sphere around the C$_{\alpha}$ atoms of each
of the residues. The side chains of the residues, which extended
beyond this first sphere, were enclosed in a second (and third)
sphere, for which the positions and radii were chosen to reproduce the
van-der-Waals surface of the peptide as close as possible. In each
residue the effective charges from the PDB structure with $|q| \leq
0.5$ e were kept. Then, the C$_{\alpha}$ spheres were connected by
springs. The diffusion coefficients of the residues were taken from Ma
etal\cite{Ma05}. Running BD simulations of this first setup we
determined the resulting mutual center-of-mass (CM) distances for each
pair of residues and compared them to the respective distances derived
from a 20 ns MD simulation of the peptide in a water box. This
reference simulation was performed with the Gromacs package
\cite{VDS05} using the OPLS all-atom force field \cite{JOR96} and the
TIP4P water model \cite{JOR83}. The MD trajectory consequently
includes the finite damping of and the hydrodynamic coupling between
the residues. As this first set of springs was not enough to reproduce
all distance distributions correctly, additional springs between the
CMs of the residues were added and their values optimized manually
until a sufficient agreement between the MD results and a standard BD
simulation was achieved. With this parametrized bead-spring-model of
the peptide we ran BD and LD simulations, each with and without
hydrodynamics. In all simulations, a single copy of the peptide was
simulated in an unbounded box. The comparison of these four different
simulation schemes shows the effects of each of the simplifications.

The calculated relaxation times of the peptide residues range from
$\tau_{rel} = 32$ fs for the small glycin (G) to 50 fs for the larger
arginine (R). For a reliable propagation we used a conservatively
estimated small time step $\Delta t = 10$ fs in all simulations.

\subsection{Center-of-Mass Diffusion}

\begin{figure}[t]
	\begin{center}
		\includegraphics[scale=0.9]{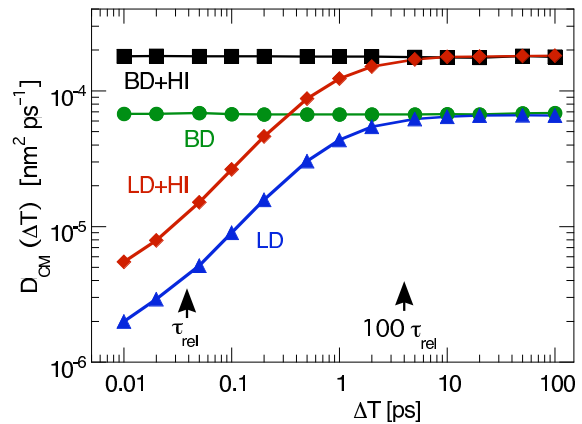}
	\end{center}
	\caption{CM diffusion coefficient $D_{CM} = \frac{\langle \Delta
	r^2\rangle}{6 \Delta T}$ of the peptide vs. the length of the
	observation time interval $\Delta T$. The arrow labeled with
	$\tau_{rel}$ indicates the typical velocity relaxation time of a
	single residue of $\approx 40$ fs. With LD the long-time value of
	$D_{CM}$ is obtained only for $\Delta T \geq 100\, \tau_{rel}$.
	With HI, $D_{CM}$ is increased by a factor of 2.7.}
	\label{fig:DCM}
\end{figure}

For the diffusion of the CM of the peptide two effects were observed.
As for polymers the CM diffusion coefficient $D_{CM}$ was increased
when HI was included. This can be well seen in figure \ref{fig:Trajs},
which compares the x-component of the CM motion of the peptide
obtained with the same sequence of random numbers for the four
different propagation schemes. All four curves show the same trends,
but the amplitudes from BD+HI and LD+HI are nearly twice as large as
without HI. In this plot one can also see that the trajectories with
LD are much smoother than with BD due to the finite relaxation time.
For both BD with and without HI the power spectrum can be well fitted
with a $1/f$ behavior, where the amplitudes with HI are about 60\%
higher. With LD, however, the amplitudes are the same in the
low-frequency range as in the corresponding BD simulations, but
fluctuations that occur on timescale shorter than a picosecond are
suppressed by about 30\%{} in this one-dimensional representation
(data not shown).

Similar conclusions can be drawn from how the CM diffusion coefficient
$D_{CM}(\Delta T)$ changes with the length of the observation time
step $\Delta T$. Figure \ref{fig:DCM} gives $D_{CM} = \langle \Delta
r^2(\Delta T)\rangle / 6\Delta T$ for the four different propagation
schemes. Here, for the same trajectory, which was simulated at a
timestep $\Delta t = 10$ fs, the observation window $\Delta T$ was
varied at which the CM displacements $\Delta \vec{r}(\Delta T)$ were
obtained. As expected, the BD simulations have no implicit time scale
and, consequently, give the same $D_{CM}$ for any $\Delta T$. The
creeping flow HI used here accelerates the diffusive motion without
introducing a timescale of its own and $D_{CM}$ is about 2.7 times
larger with hydrodynamics than without. For long observation
intervals $\Delta T > 10$ ps, the LD propagation reproduces the
respective BD values for $D_{CM}$, whereas in the ballistic short time
regime $D_{CM}$ increases linearly with $\Delta T$. The relaxation
times of the individual residues are in the range $\tau_{rel} = 32
\ldots 50$ fs, which is indicated in figure \ref{fig:DCM}. However, it
takes about two orders of magnitude longer, until $D_{CM}$ from the LD
propagation reaches the timescale-free BD value.

\subsection{Orientational Relaxation}

\begin{figure}[t]
	\begin{center}
		\includegraphics[scale=0.9]{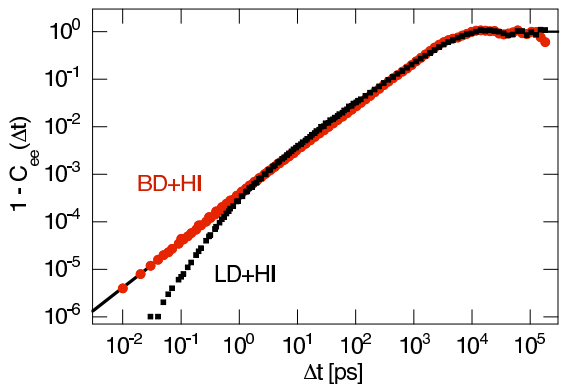}
	\end{center}
	\caption{Orientational relaxation of the end-to-end vector
	$R_{ee}$ as seen from the autocorrelation $C_{ee}$ from BD+HI (red
	points) and LD+HI simulations (black squares). The solid line
	indicates the fit to a stretched exponential as explained in the
	text.}
	\label{fig:Relax}
\end{figure}

As shown above, the inclusion of HI accelerates the collective CM
diffusion. This is accompanied by a slowdown of the internal dynamics.
For bead-spring polymers this can, e.g., be seen in a prolonged
correlation of the end-to-end vector $\vec{R}_{ee}$ pointing from the
first to the last monomer. The peptide used here is too short and not
flexible enough to show pronounced differences in the relaxation of
$\vec{R}_{ee}$ whether HI are included or not. Nevertheless, the
effects of the finite damping can be seen in the autocorrelation
$C_{ee}(\Delta t) = \langle \hat{R}_{ee}(t) \hat{R}_{ee}(t+\Delta
t)\rangle$, which is plotted in figure \ref{fig:Relax} as
$1-C_{ee}(\Delta t)$. For both BD and BD+HI, $C_{ee}$ fits well to a
stretched exponential
\be{eq:StrechedExp}
	C_{ee}(\Delta t) = 
		\exp \left[-\left(\Delta t / \tau\right)^\alpha\right]
\ee
with $\tau$ = 4.1 ns and $\alpha = 0.95$. With the LD
propagation, the dynamics on timescales shorter than a few picoseconds
is slowed down both with and without HI. Correspondingly, in figure
\ref{fig:Relax}, $1 - C_{ee}$ from the Langevin simulation is smaller
than with the overdamped Brownian propagation for short time
intervals, while for longer intervals the same exponential decay is
observed.


\subsection{Internal Dynamics}

\begin{figure}[t]
	\begin{center}
		\includegraphics[scale=0.9]{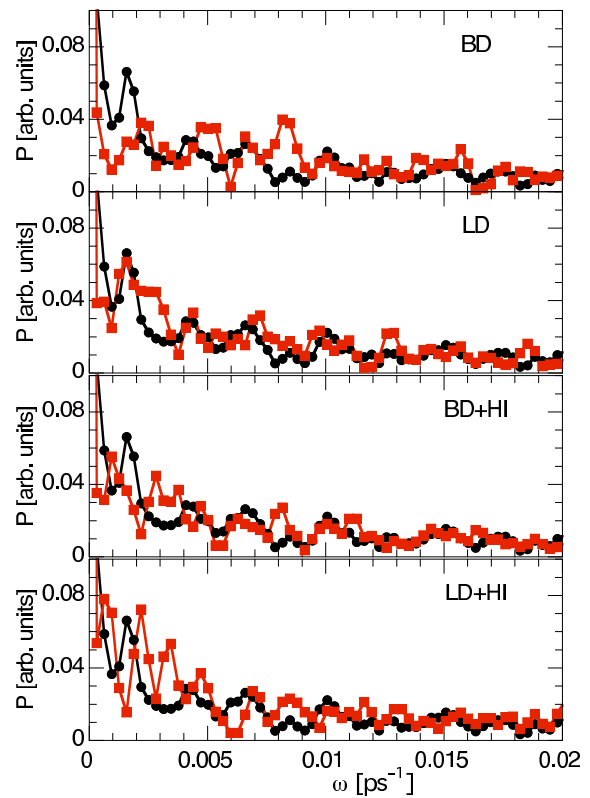}
	\end{center}
	\caption{Comparison of the power spectra for the dynamics of the
	CM distance between the first and the last residue with the
	different propagation schemes. In all four panels the black data
	(points) are from the MD simulation and the red (squares) from the
	respective coarse grained method.}
	\label{fig:Frequ}
\end{figure}

The bonds of the coarse-grained representation of the peptide had been
parametrized against the stationary distance distributions without
considering the actual dynamics, which are also influenced by the
residue masses and the damping. Consequently, the dynamics will have
different temporal signatures with the different propagation schemes.
Figure \ref{fig:Frequ} shows the power spectrum of the distance between
the first and the last bond for the four integration schemes in
comparison to the power spectrum extracted from the MD trajectory.
These power spectra are typical for the distances between residues
located on opposite ends of the peptide.

As seen in figure \ref{fig:Frequ}, the first four peaks between 0.002
ps$^{-1}$ and 0.01 ps$^{-1}$ are present in the LD result and can be
recognized in the LD+HI spectrum. BD and BD+HI are damped too fast.
Consequently, no real peak structure can be identified above the
strongly fluctuating background. For the other distances between
residues on opposite ends of the peptide, a similar trend is found.

These spectra now could be further improved by optimizing the
effective masses of the residues. However, already these results with
the raw textbook-masses demonstrate that the short-time dynamics on
the picosecond timescale are better reproduced with the finitely
damped Langevin algorithm.


\subsection{Numerical Stability and Runtime}

The first test had been to compare the distance distributions between
the CMs of the residues for all four simulation schemes. As expected,
for these static measures no differences beyond the statistical
uncertainties were found. Getting the same positions and widths with
LD as with BD shows that the velocity relaxation occurs faster than
one oscillation period in the interparticle potentials even for the
rather hard springs that had to be used along the peptide backbone
between the C$_{\alpha}$ atoms. Otherwise one would expect a widening
of the distance distributions and, in the extreme case, a bimodal
density as in a classical weakly damped oscillator.

\begin{figure}[t]
	\begin{center}
		\includegraphics[scale=0.9]{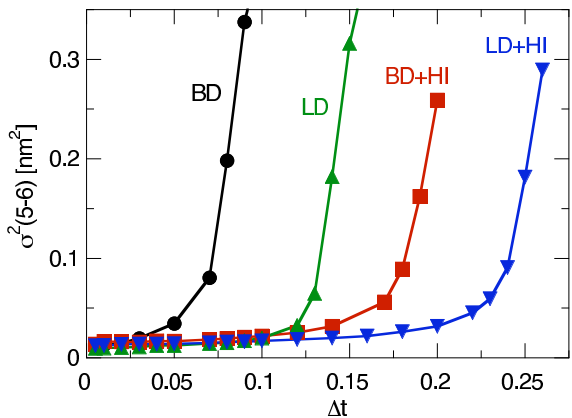}
	\end{center}
	\caption{Variance $\sigma^2$ of the distance distribution of the
	bond between residues 5 and 6 vs. the simulation timestep $\Delta
	t$ for the four propagation schemes.}
	\label{fig:Breite56}
\end{figure}

The stability of each integration scheme was then tested via the
variances of the distance distributions for the hard springs, i.e.,
those distributions which have a small variance. Figure
\ref{fig:Breite56} shows the representative behavior for the distance
between the fifth and the sixth residue, which are both part of the
stiff PP2 helix in the central part of the peptide. With BD alone,
timesteps up to 0.05 ps would have been okay, whereas the viscous
damping due to HI allows for three times longer timesteps. The same
trend is observed for LD, which itself is already more stable than BD
because of the damping at high frequencies.

The respective runtimes for 100000 timesteps on a 2.8 GHz Pentium 4
CPU were 18.9 s for BD, 24.2 s for BD+HI, 19.5 s for LD, and 24.5 s
for LD+HI. As explained above, keeping track of the velocity in the LD
propagation only adds a minimal cost which already pays off when only
the numerical stability is taken into account and considerations about
the regime of damping are ignored. In this small peptide with its 35
effective charges and 38 bonds, the hydrodynamic interactions between
the eleven residues only add a manageable 25\% to the total runtime, 
whereby the direct interactions and the hydrodynamic coupling have the 
same $\mathcal{O}(N^2)$ runtime scaling.

\section{Summary and Conclusions}

In this publication we investigated the effects of a finite damping
and of hydrodynamic interactions in Brownian dynamics simulations of a
small biological peptide.

Integrating Newton's equations of motion with a continuous viscous
solvent over one timestep, we arrived at a Langevin type integration
scheme which allows for the inclusion of our recently introduced
efficient approximation for the hydrodynamic coupling \cite{GEY09}.
With a fast damping this Langevin scheme reduces to the conventional
BD algorithm of Ermak and McCammon \cite{ERM78}. Due to the analytic
integration for one timestep the Langevin propagator is about as fast
numerically as the simple BD integration, but does not have the
conceptual limitation that the timestep $\Delta t$ has to be much
longer than the velocity relaxation time $\tau_{rel}$. This issue
arises especially in simulations of proteins of different sizes,
because there $\Delta t$ may be too close to $\tau_{rel}$ for the
larger proteins, while for the smaller, faster proteins $\Delta t$ is
already too long for a stable propagation. With the LD propagator, the
length of $\Delta t$ is only limited by numerical accuracy and
stability considerations. Actually, our test showed that only when
$\Delta t$ is at least two orders of magnitude longer than
$\tau_{rel}$, the finite damping may safely be neglected. For
practical applications such a long integration timestep is nearly
never feasible.

Comparing the dynamics of a small peptide from our coarse grained BD
and LD simulations to an atomistic molecular dynamics trajectory, we
found that the distance distributions between the residues of the
peptide can be reproduced with a bead-spring model with either
propagation technique and the same force constants. However, when the
dynamics of the relative motions are compared, LD with its more
realistic finite damping gives an overall better agreement of the
internal dynamics.

For the short and rather stiff peptide investigated here, the
influence of hydrodynamic interactions on the internal motion was
small. The center-of-mass motion, however, became faster by nearly a
factor of three. Compared to a plain BD simulation, the slightly
delayed orientational motion with the LD scheme together with the
faster translation due to the HI may change the microscopic dynamics
during, for example, the association of two proteins or the binding of
a small peptide to a larger protein. At close encounter the smoothed
motion with LD may then further modify the obtained association and
dissociation rates towards a more stable bound state. From a
conceptual point of view we therefore recommend to include
hydrodynamics and to use a Langevin propagation scheme for coarse
grained protein simulations.

From a numerical point of view, we find no difference in runtime for
BD vs. LD and a moderate increase in runtime of about 30\% for adding
our approximate hydrodynamics with its $\mathcal{O}(N^2)$ runtime
scaling. The coarse grained model of the peptide had about six times
as many effective charges on the individual residues than there were
effective spheres for the hydrodynamics. For peptide or protein models
with a similar ratio of the numbers of the simple Coulomb compared to
the more expensive hydrodynamic interactions, the runtime cost incurred
by HI will be similar.

The finitely damped LD propagation was also more stable numerically,
i.e., one could use longer integration timesteps than with the
overdamped BD algorithm. The numerical stability of both the BD and
the LD schemes further increased with the viscous damping due to the
HI. Actually, the longer timesteps that could be used with LD and HI
more than compensated for the increased runtime.

Here we used a simple peptide to demonstrate that coarse grained
simulation methods need not stop at the level of Einstein's
formulation of Brownian diffusion but that they can also be used very
efficiently for much smaller systems, which are usually considered the
realm of atomistic simulations or partially coarsened methods, where a
few atoms each are summarized into one pseudo atom \cite{MAR04,GUI06}.
The main difference, which makes fine-grained BD and LD simulations so
much faster than the coarse grained versions of all-atom descriptions
is that the very many solvent molecules do not have to be treated
explicitly.

Summarizing we could show that the LD+HI simulation scheme, which
includes more of the mechanically induced solvent effects than the
conventional BD algorithm, can be formulated and implemented
efficiently. Consequently, we are looking forward to applying the
methods presented here to larger and more realistic systems in the
near future. Potential applications will be the investigation of
association funnels of peptide-protein encounter \cite{SPA06, AHM08},
the folding of coarse grained proteins \cite{FRE09}, and dense
many-particle scenarios like the cytosol inside a biological cell.

The LD and HI algorithms presented here have been implemented in a
general purpose multi-particle BD/LD simulation library, which will be
available from the corresponding author free of charge for academic
use.

\section{Acknowledgements}

We thank Ahmad Mazen for the post-processed atomic coordinates and the
molecular dynamics trajectory of the simulated peptide. This work was
funded by the Deutsche Forschungsgemeinschaft through the
Graduiertenkolleg 1276/1.



\begin{thebibliography}{99}

	\bibitem{PHI88} R. J. Phillips, J. F. Brady and G. Bossis,
	Phys. Fluids \textbf{31} 3462 (1988)
	
	\bibitem{LAD90} A. J. C. Ladd,
	J. Chem. Phys. \textbf{93} 3484 (1990)
	
	\bibitem{GOR04} C. Gorba, T. Geyer, and V. Helms,
	J. Chem. Phys. \textbf{121}, 457 (2004)
	
	\bibitem{LIU03} B. Liu and B. D\"unweg,
	J. Chem. Phys. \textbf{118}, 8061 (2003)

	\bibitem{BEA00b} D. A. Beard and T. Schlick,
	J. Chem. Phys. \textbf{112} 7323 (2000)
	
	\bibitem{SZY07} P. Szymczak and M. Cieplak,
	J. Chem. Phys. \textbf{127}, 155106 (2007)
	
	\bibitem{EMP08} A. Emperador, O. Carrillo, M. Rueda and M. Orozco,
	Biophys. J. \textbf{95} 2127 (2008)
	
	\bibitem{MIN05} D. D. J. Minh, J. M. Bui, C. Chang, T. Jain, J. 
	M. J. Swanson and A. J. McCammon,
	Biophys. J. \textbf{89} L25 (2005)
	
	\bibitem{GAB02} R. R. Gabdoulline and R. C. Wade,
	Curr. Op. Struc. Biol. \textbf{12}, 204 (2002)

	\bibitem{SPA06} A. Spaar, C. Dammer, R. R. Gabdoulline, R. C. 
	Wade, and V. Helms,
	Biophys. J. \textbf{90}, 1913 (2006)
	
	\bibitem{DLU08} M. Dlugosz, J. A. Antosiewicz, and J. Trylska,
	J. Chem. Theory Comput. \textbf{4}, 549 (2008)
	
	\bibitem{EIN05} A. Einstein,
	Ann. d. Physik \textbf{17}, 549 (1905)

	\bibitem{DHO96} J. K. G. Dhont, 
	\emph{An Introduction to Dynamics of Colloids}
	(Elsevier, Amsterdam, 1996)
	
	\bibitem{FRE09} T. Frembgen-Kesner and A. H. Elcock
	J. Chem. Theory Comput. \textbf{5} 242 (2009)

	\bibitem{ERM78} D. L. Ermak and J. A. McCammon,
	J. Chem. Phys. \textbf{69}, 1352 (1978)

	\bibitem{FIX81} M. Fixman,
	Macromolecules \textbf{14} 1710 (1981)

	\bibitem{GEY09} T. Geyer and U. Winter, 
	J. Chem. Phys. \textbf{130} 114905 (2009)
	
	\bibitem{BEA00a} D. A. Beard and T. Schlick,
	J. Chem. Phys. \textbf{112} 7313 (2000)

	\bibitem{KIR48} J. G. Kirkwood and J. Riseman,
	J. Chem. Phys. \textbf{16}, 565 (1948)
	
	\bibitem{ROT69} J. Rotne and S. Prager,
	J. Chem. Phys. \textbf{50}, 4831 (1969)
	
	\bibitem{YAM70} H. Yamakawa,
	J. Chem. Phys. \textbf{53}, 436 (1970)
	
	\bibitem{FAX22} H. Faxen,
	Ann. d. Phys. \textbf{373} 89 (1922)
	
	\bibitem{AHM08} M. Ahmad, W. Gu and V. Helms, 
	Angew. Chem. Int. Ed. \textbf{47} 7626 (2008)
	
	\bibitem{ALB02} B. Alberts, A. Johnson, J. Lewis, M. Raff, K. 
	Roberts and P. Walter,
	\emph{Molecular Biology of the Cell, 4th Ed.}
	(Garland Science, New York, Amsterdam, 2002)
	
	\bibitem{VDS05} D. Van Der Spoel, E. Lindahl, B. Hess, G. 
	Groenhof, A. E. Mark and H. J. Berendsen, 
	J. Comput. Chem. \textbf{26}, 1701 (2005)
	
	\bibitem{JOR96} W. L. Jorgensen, D. S. Maxwell and J. Tirado-Rives, 
	J. Am. Chem. Soc. \textbf{118}, 11225 (1996)
	
	\bibitem{JOR83} W. L. Jorgensen, J. Chandrasekhar, J. D. Madura, 
	R. W. Impey and M. L. Klein,
	J. Chem. Phys. \textbf{79}, 926 (1983)
	
	\bibitem{Ma05} Y. Ma, C. Zhu, P. Ma and K. T. Yu,
	J. Chem. End. Data \textbf{50}, 1192 (2005)

	\bibitem{MAR04} S. J. Marrink, A. H. de Vries and A. E. Mark,
	J. Phys. Chem. B \textbf{108} 750 (2004)
	
	\bibitem{GUI06} G. Guigas and M. Weiss,
	Biophys. J. \textbf{91} 2393 (2006)
	
\end{thebibliography}
\end{document}